# Direct imaging of the coexistence of ferromagnetism and superconductivity at the LaAlO$_3$/SrTiO$_3$ interface

J. A. Bert,[1] B. Kalisky,[1] C. Bell,[2] M. Kim,[2] Y. Hikita,[2] H. Y. Hwang,[1,2] and K. A. Moler[1]

1. Stanford Institute for Materials and Energy Science, Stanford University, Stanford, California 94305, USA
2. Department of Advanced Materials Science, University of Tokyo, Kashiwa, Chiba 277-8561, Japan

**LaAlO$_3$ and SrTiO$_3$ are insulating, nonmagnetic oxides, yet the interface between them exhibits a two-dimensional electron system with high electron mobility,[1] superconductivity at low temperatures,[2-6] and electric-field-tuned metal-insulator and superconductor-insulator phase transitions.[3,6-8] Bulk magnetization and magnetoresistance measurements also suggest some form of magnetism depending on preparation conditions[5,9-11] and suggest a tendency towards nanoscale electronic phase separation.[10] Here we use local imaging of the magnetization and magnetic susceptibility to directly observe a landscape of ferromagnetism, paramagnetism, and superconductivity. We find submicron patches of ferromagnetism in a uniform background of paramagnetism, with a nonuniform, weak diamagnetic superconducting susceptibility at low temperature. These results demonstrate the existence of nanoscale phase separation as suggested by theoretical predictions based on nearly degenerate interface sub-bands associated with the Ti orbitals.[12,13] The magnitude and temperature dependence of the paramagnetic response suggests that the vast majority of the electrons at the interface are localized, and do not contribute to transport measurements.[3,6,7] In addition to the implications for magnetism, the existence of a 2D superconductor at an interface with highly broken inversion symmetry and a ferromagnetic landscape in the background suggests the potential for exotic superconducting phenomena.**

Coexistence of ferromagnetism and superconductivity in nature is rare.[14-18] The LaAlO$_3$/SrTiO$_3$ interface is a new system for studying this coexistence. LaAlO$_3$ (LAO) and SrTiO$_3$ (STO) are both perovskite band insulators with no magnetic order in their bulk form. For LAO grown on the TiO$_2$ terminated STO substrate, a high mobility electron gas was observed at the interface.[1] Electronic reconstruction, driven by the polar/nonpolar interface, moves charge from the LAO layers across the interface into the STO causing an effective electronic doping responsible for the observed conductivity.[1] The interplay of this effect with oxygen vacancies and structural changes,[19] and the relative contribution of these three effects to the carrier concentration, remains a subject of debate. Significant variability in the physical properties in similar samples indicates that the ground state of this interface system is sensitive to small changes in growth conditions. Superconductivity[2-5] and features interpreted as interface magnetism[5,9,10] have been independently observed at the LAO/STO interface via transport and bulk magnetization measurements. One recent study inferred the existence of both ferromagnetism and superconductivity in the same sample from hysteresis in magnetoresistance transport measurements.[5]

We use a scanning superconducting quantum interference device (SQUID) with micron-scale spatial resolution to image three samples down to 20 mK (See methods). Our SQUID sensor can concurrently measure the static magnetic fields generated by the sample (magnetometry) and the susceptibility of the sample to a small locally applied ac magnetic field (susceptometry). Fig 1 a&b show magnetometry and susceptometry images of an LAO/STO interface. The ferromagnetic landscape appears as many static spatially separated dipoles that show no temperature dependence over the measured temperature range. The superconductivity is spatially inhomogeneous and weak, with a critical temperature $T_c$ = 100 mK (Fig 1c), above which a temperature-dependent paramagnetic response is apparent (Fig 1c inset). In contrast, a delta-doped STO sample[20] has relatively uniform 2D superconductivity, no magnetic order, and no apparent paramagnetic response above $T_c$ (Fig 1d,e,&f), although the expected paramagnetic signal at $T_c$ is close to our noise floor.

The diamagnetic susceptometry from the LAO/STO interface is an order of magnitude smaller than that of the delta-doped $SrTiO_3$ or $(Ba_{0.9}Nb_{0.1}CuO_{2+x})_m/(CaCuO_2)_n$, another two-dimensional superconductor.[21] The susceptometry signal is generated by superconducting electrons which screen the local applied field and is related to the local density of electrons in the superconducting condensate. The superfluid density is usually quantified by the magnetic penetration depth, $\lambda$.[22,23] In a 2D superconductor with thickness $d<<\lambda$, the screening currents are confined in the vertical direction which generates a modified penetration depth known as the Pearl length, $\Lambda=2\lambda^2/d$. The low temperature Pearl length in the δ-doped STO sample was 650 μm based on fits to formulas for the height dependence of the susceptometry from references [23,24]. This formula should not quantitatively describe the data for the LAO/STO interface due to the lateral inhomogeneities, but the susceptibility signal from a uniform 2D superconductor scales as $1/\Lambda$ for large $\Lambda$, implying an ~8 mm Pearl length in the LAO/STO.

The $T_c$ variation between two measurement positions on the LAO/STO sample (Fig 1c) is about 10%. However, the lateral variation of the susceptometry is large: 84% of the total response, compared to just 12% in the doped STO, and less than 1% in most bulk superconductors.[25] The largely inhomogeneous superconducting and ferromagnetic response may suggest proximity to a first order phase transition. Although both magnetism and superconductivity are present at the interface in the LAO/STO sample, Fig 1 a&b do not show a direct correlation between the inhomogeneity of the superconducting state and the distribution of magnetic regions.

The ferromagnetism appears as magnetic dipoles in Fig 1a and Fig 2a, mostly separated from each other by microns, with many additional dipoles that do not show up visually in these images but are still above our noise threshold (Fig 2a insets). We analyzed six 70x80 micron high resolution magnetometry scans, including the one shown in Fig 2a, finding 144 dipoles above our noise floor and fitting each one to a point dipole model to determine its total moment and orientation (Fig 2b-d). The histogram of the dipole moments shows a clear exponential distribution of dipole moments with a few large (~$1 \times 10^8 \mu_B$) dipoles and substantially more smaller dipoles down to the limit of our noise. This trend suggests that there are even more dipoles with moments below the sensitivity of our SQUID.

Most of the dipoles lie in plane, as expected from the shape anisotropy of the interface, with apparently randomly distributed azimuthal angles indicating no alignment or net magnetization. This observation is consistent with cooling the sample in zero field. The point dipole approximation is not as good for some dipoles, particularly the ones with the largest moments, indicating that they are not point-like but are instead ferromagnetic patches that extend over an area comparable to the SQUID's 3 µm pick-up loop.

The dipoles were stable throughout the duration of the cooldown (about 1 month) and were insensitive to temperature changes from 20 mK through the superconducting critical temperature and up to 4.2 K. Additional SQUID measurements in a separate variable temperature cryostat showed that the dipole size and orientation remained unchanged between 4.2 K and our maximum measurement temperature of 60 K. In addition, we measured a second sample with 10 uc of LAO grown on a $TiO_2$ terminated surface that had patterned Hall bars (Methods). This second 10 uc LAO/STO sample had many fewer dipoles – none in some regions. The variability in the size of the moment may be related to the variability of physical properties in nominally identical samples in this system.

We did not observe dipoles in the magnetometry signal on the doped STO sample (Fig 1d). Since both the doped STO sample and the LAO/STO samples use the same commercially available STO substrates, the absence of dipoles on the doped STO sample rules out magnetic impurities in the substrate.

In addition to the ferromagnetic order, the two LAO/STO samples measured at low temperature show paramagnetism above the superconducting critical temperature $T_c$ (Fig 1c inset, Fig 3). In the case of the patterned LAO/STO sample, which did not show many ferromagnetic dipoles, we observe regions where no superconductivity appears and the paramagnetism remains down to the lowest measured temperatures. The paramagnetic signal decreases with increasing temperature suggesting a Curie law. The $1/T$ dependence and the paramagnetic sign indicate that the susceptibility signal originates from localized spins.

We can estimate the electron density associated with the ferromagnetic, diamagnetic, and paramagnetic signals. We determine the number of ferromagnetic electrons by adding the moments of all the dipoles in the histogram yielding $7.3 \pm 3.4 \times 10^{12} \mu_B/cm^2$. This estimate is a lower bound, because any dipoles that are below the sensitivity of our sensor or whose moments canceled due to the random distribution of alignments have not been included in this total. We use the Pearl length to find the density of superconducting electrons, $n_s = 2m^*/\mu_0 e^2 \Lambda$, where $e$ is the elementary charge and $\mu_0$ is the permeability of free space. Using $m^* = 1.45\,m_e$ [26] we find $n_s \approx 1\times10^{12}\,\text{cm}^{-2}$. We quantify the paramagnetic signal by using an appropriate model for our sensor to convert our measured susceptibility, $\phi$, to the dimensionless susceptibility, $\chi$, for a layer of spins in a thickness $d$. Using $\chi d = 22\,\mu\text{m}\cdot\text{mA}/\Phi_0\cdot\phi$ [27] and comparing $\chi$ to the Curie expression, $\chi=\mu_0 n_{3D}(g\mu_B)^2 J(J+1)/3k_B T$ with $g=2$ and $J=1/2$, yields a 2D spin density of $4.4\times10^{14}\,\text{cm}^{-2}$, with large error bars due to uncertainty in the geometrical parameters. We compare our estimates with the electron densities predicted by the polar catastrophe, $3\times10^{14}\,\text{cm}^{-2}$, and seen in hall measurements,[3,6,7] $1\text{-}4\times10^{13}\,\text{cm}^{-2}$ (Table 1). The densities of magnetic and superconducting electrons are respectively one and two orders of magnitude lower than the polar catastrophe density, but the paramagnetic spin density shows surprising agreement within error.

Density functional calculations of the electronic structure in LAO/STO predict the presence of multiple nearly degenerate subbands that result in separate charge carriers.[12] Magnetism was also predicted at the n-type LAO/STO interface from alignment of additional electrons in the Ti orbitals.[13] Transport measurements, which probe delocalized electrons, have measured electron densities significantly lower than predictions from the polar catastrophe. Our measurements indicate that those missing electrons may be present but localized, and contribute to the magnetic signal.

The observation of ferromagnetism and superconductivity at the LAO/STO interface opens exciting possibilities for studying the interplay of these normally incompatible states. Tuning the carriers with a gate voltage may add even more richness to the system, by coincidently studying how adding or removing carriers affect the superconducting, ferromagnetic and paramagnetic signals.

**Methods Summary**

The two LAO/STO samples used in the low temperature study were prepared by growing 10 unit cells of LaAlO$_3$ on commercial TiO$_2$ terminated {001} STO substrates. The patterned sample had an AlO$_x$ hard mask which defined hall bars. The LaAlO$_3$ was deposited at 800°C with an oxygen partial pressure of $10^{-5}$ mbar, after a pre-anneal at 950°C with an oxygen partial pressure of $5\times10^{-6}$ mbar for 30 minutes. The samples were cooled to 600°C and annealed in a high pressure oxygen environment (0.4 bar) for one hour.[6]

A δ-doped STO sample was also studied at low temperatures. It was grown in an atmosphere of less than $10^{-8}$ torr oxygen at 1,200 °C. Nb dopants were confined to a 5.9nm layer and additional 100 nm cap and buffer layers of STO were grown above and below the doped region. The sample was annealed in situ at 900 °C under an oxygen partial pressure of $10^{-2}$ torr for 30 minutes.[20]

Measurements were done by scanning SQUID in a dilution refrigerator.[28,29] The SQUID has a 3 µm pick-up loop, centered in a single turn field coil. Static magnetism (magnetometry) in the sample is probed by recording the flux through the SQUID pick-up loop as a function of position. Applying an ac current in the field coil produces a local magnetic field. The local susceptibility (susceptometry) of the sample to the applied field is detected by the pick-up loop in a lock-in measurement.

**Acknowledgments**

We thank Martin Huber for assistance in SQUID design and fabrication. This work was primarily supported by the U.S. DOE, Division of Materials Sciences, under Award No. DE-AC02-76SF00515. B.K. acknowledges support from FENA. H.Y.H. acknowledges support by the Department of Energy, Office of Basic Energy Sciences, Division of Materials Sciences and Engineering, under contract DE-AC02-76SF00515.

**Author Contributions**

SQUID measurements: J.A.B and B.K. Analysis: J.A.B and B.K. with ideas developed with H.Y.H. and K.A.M. Sample growth: C.B., M.K., Y.H., and H.Y.H. Manuscript preparation: J.A.B., H.Y.H and K.A.M., with input from all co-authors.

**Author Information**

Reprints and permissions information is available at www.nature.com/reprints. Correspondence and requests for materials should be addressed to J.A.B (jbert@stanford.edu).

|  | Electron Density |
|---|---|
| Polar Catastrophe | $3.2 \times 10^{14}$ cm$^{-2}$ |
| Paramagnetic Spins | $1-5 \times 10^{14}$ cm$^{-2}$ |
| Hall Effect[3,6,7] | $1-4 \times 10^{13}$ cm$^{-2}$ |
| Dipole Moment | $7.3 \pm 3.4 \times 10^{12}$ cm$^{-2}$ |
| Superfluid Density | $0.9-1.7 \times 10^{12}$ cm$^{-2}$ |

**Table 1|Table of electron densities**

Extracted from hall measurements, measurement of the ferromagnetic, superconducting and paramagnetic signals.

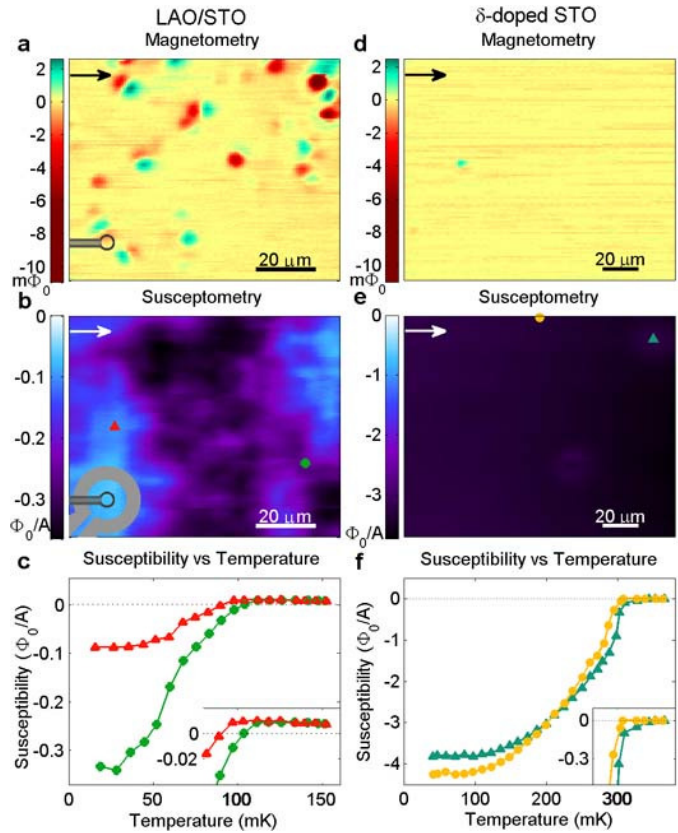

**Figure 1| Comparison of SQUID images on LAO/STO and δ-doped STO samples.**

**a**, LAO/STO magnetometry image mapping the ferromagnetic order. Inset, scale image of the SQUID pick-up loop used to sense magnetic flux. **b**, LAO/STO susceptometry image mapping the superfluid density at 40 mK. Inset, scale image of the SQUID pick-up loop and field coil. **c**, The temperature dependence of the susceptibility taken at the two positions indicated in **b**. **d**, δ-doped STO magnetometry image showing no ferromagnetic order. **e**, δ-doped STO susceptometry image mapping the superfluid density at 82 mK. **f**, The temperature dependence of the susceptibility taken at the two positions indicated in **e**. The arrow on each scan shows the scan fast axis and the SQUID orientation.

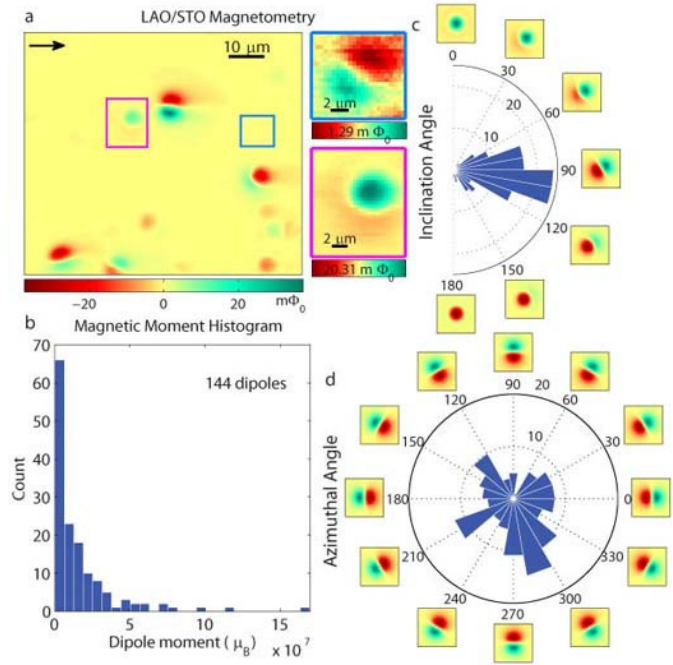

**Figure 2| Analysis of the dipole distribution.**

**a**, Magnetometry scan showing ferromagnetic dipoles. The arrow shows the scan fast axis and the SQUID orientation. Insets: Individual dipoles from the areas indicated in the larger image. **b-d**, Histograms of the moment and orientation of 144 dipoles taken from six large area scans similar to the one show in panel **a**. **b**, The magnetic moment of each dipole in Bohr magnetons, $\mu_B$. **c**, The inclination angle from the normal to the sample surface (an inclination angle of 90 degrees mean the dipole lies in the plane of the interface). **d**, The azimuthal angle with respect to the scan's x-axis.

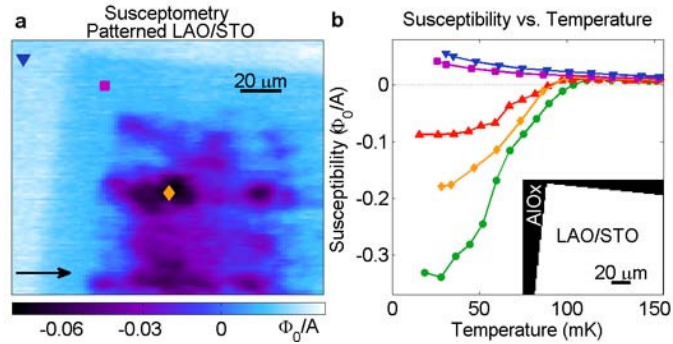

**Figure 3| Paramagnetic signal on patterned LAO/STO sample.**

**a**, Susceptometry scan on the patterned sample at 87 mK. A suppression of the diamagnetic susceptibility is visible near the edge of the pattern. The susceptibility response in this area has a paramagnetic sign as indicated in the susceptibility vs. temperature plot. The arrow indicates the scan fast axis and the SQUID orientation. **b**, Susceptibility vs. temperature data from 3 positions on the patterned LAO/STO shown in panel **a**. The red triangles and green circles are data reproduced from the unpatterned LAO/STO sample show in Fig 1a. Inset: The outline of the AlO$_x$ patterning associated with the susceptometry scan.